\documentclass[aps,preprint]{revtex4}
\usepackage{graphicx}
\usepackage{times}

\begin{document}

\title{Lattice Boltzmann Algorithm for three-dimensional liquid crystal hydrodynamics}
\author{C. Denniston$^1$, D. Marenduzzo$^2$, E. Orlandini$^3$, and J.M. Yeomans$^2$}
\affiliation{
$^1$Department of Applied Mathematics, The University of Western 
Ontario, London, Ontario N6A 5B8, Canada\\
$^2$Department of Physics, Theoretical Physics, 1 Keble Road, Oxford, OX1 3NP, England\\
$^3$INFM, Dipartimento di Fisica, Universita' di Padova, Via Marzolo 8, 35131 Padova, Italy}

\begin{abstract}
We describe a lattice Boltzmann algorithm to simulate liquid crystal
hydrodynamics in three dimensions. 
The equations of motion are written in terms of a
tensor order parameter. This allows both the isotropic and the nematic
phases to be considered. Backflow effects and the hydrodynamics of
topological defects are naturally included in the simulations, as are
viscoelastic effects such as shear-thinning and shear-banding.
We describe the implementation of velocity boundary 
conditions and show that the algorithm can be used to describe optical bounce in twisted nematic devices and secondary flow in sheared nematics with an imposed twist.
\end{abstract}

\maketitle

\renewcommand{\theequation}{\thesection.\arabic{equation}}
\renewcommand{\baselinestretch}{1.2} \def\xbar{1/x} \def\ra{\rangle}
\def\la{\langle} \def\Pr{{\it Proof: }} \def\qed{$\Box$} \def\cF{{\cal
F}} \def\cFo{{\cal F}^o} \def\pan{\par\noindent}
\def\pa{\partial_{\alpha}} \def\pb{\partial_{\beta}}
\def\pe{\partial_{\eta}} \def\pt{\partial_t
} \def\eia{e_{i\alpha}}
\def\eib{e_{i\beta}} \def\eie{e_{i\eta}} \def\geq{g_i^{eq}}
\def\feq{f_i^{eq}} \def\dt{\Delta t} \def\ua{u_{\alpha}}
\def\ue{u_{\eta}} \def\Tea{T_{\eta\alpha}} \def\Pea{P_{\eta\alpha}}
\def\Qab{Q_{\alpha\beta}}\def\Pab{-\sigma_{\alpha\beta}}
\def\tauab{\tau_{\alpha\beta}}\def\Hab{H_{\alpha\beta}}

\section{Introduction}
Liquid crystals are fluids, typically comprising long thin molecules. Subtle 
energy-entropy balances can cause the molecules to align to form a variety 
of ordered states. For example, in nematic liquid crystals, the molecules 
tend to align parallel giving a state with long-range orientational order. 
Liquid crystals exhibit both an elastic and a viscous response to an 
external stress. Coupling between the director and velocity fields leads to 
strongly non-Newtonian flow behaviour such as shear-banding and molecular 
tumbling under an applied shear.

Given both the rich rheological behaviour of liquid crystals and their importance in optical devices there is need to develop a robust numerical method to allow us to explore their dynamics. The hydrodynamic equations of motion are complex and, although much interesting analytic work has been carried out
(see de Gennes \& Prost 1993, Beris \& Edwards 1994, and the 
recent review Rey \& Denn 2002), there are many interesting questions that 
cannot be answered analytically. Therefore, in this paper, we describe a 
lattice Boltzmann approach to liquid crystal hydrodynamics in three 
dimensions. This generalizes a two dimensional algorithm which was 
proposed by Denniston et al. (2000). For a review of the lattice Boltzmann 
algorithm see Chen \& Doolen (1998), S. Succi (2000), Wolf-Gladrow (2000),
R. Benzi et al. (1992).

We first summarize the Beris-Edwards equations of motion for nematic liquid 
crystals (Beris {\it et al.}, 1990,1994). 
The lattice Boltzmann algorithm is described in some detail. We pay specific 
attention to boundary conditions which can be more cumbersome in three 
dimensions than in two. We then present two examples, switching in a twisted 
nematic device and shear flow in a twisted nematic, to illustrate the 
applicability of the method.

\section{Modelling liquid crystal hydrodynamics}

\subsection{Equations of motion}

Liquid crystals are described in terms of a local tensor order parameter 
${\bf Q}$,  that is related to the direction of individual molecules $\hat m$
by $Q_{\alpha\beta}=\langle {\hat m}_\alpha {\hat m}_\beta-\frac{1}{3}
\delta_{\alpha\beta}\rangle$, where the angular brackets denote a 
coarse-grained average.
Greek indices will be used to represent Cartesian components of vectors and 
tensors, and the usual summation over repeated indices will be assumed.  
${\bf Q}$ is a traceless symmetric tensor.  Its largest eigenvalue, 
$\frac{2}{3}q$, $0<q<1$, describes the magnitude of order along its principle 
eigenvector $\hat {\bf n}$, referred to as the director.

The order parameter evolves according to the equation 
(Beris {\it et al.}, 1990,1994)
\begin{equation}
(\partial_t+{\vec u}\cdot{\bf \nabla}){\bf Q}-{\bf S}({\bf W},{\bf
  Q})= \Gamma {\bf H}
\label{Qevolution}
\end{equation}
where $\Gamma$ is a collective rotational diffusion constant and
\begin{eqnarray}
{\bf S}({\bf W},{\bf Q})
&=&(\xi{\bf D}+{\bf \Omega})({\bf Q}+{\bf I}/3)+({\bf Q}+
{\bf I}/3)(\xi{\bf D}-{\bf \Omega})\nonumber\\
& & -2\xi({\bf Q}+{\bf I}/3){\mbox{Tr}}({\bf Q}{\bf W})
\end{eqnarray}
where ${\bf D}=({\bf W}+{\bf W}^T)/2$ and
${\bf \Omega}=({\bf W}-{\bf W}^T)/2$
are the symmetric part and the anti-symmetric part respectively of the
velocity gradient tensor $W_{\alpha\beta}=\partial_\beta u_\alpha$.  The 
mixture of upper and lower convective derivatives is governed by the constant
$\xi$, which depends on the molecular details of a given liquid crystal.

The term on the right-hand side of Eq.(\ref{Qevolution}) describes the 
relaxation of the order parameter towards the minimum of the free energy.
The driving motion is provided by the molecular field ${\bf H}$, which is
related to the variational derivative of the free energy ${\mathcal F}$ by
\begin{equation}
{\bf H}= -\frac{\delta {\cal F}}{\delta Q}+({\bf
    I}/3) Tr\frac{\delta {\cal F}}{\delta Q}
\end{equation}
The symmetry and zero trace of ${\bf Q}$ (and ${\bf H}$) is exploited for 
simplification.

The fluid has density $\rho$ and obeys both the continuity equation
\begin{equation}
\pt \rho + \pa \rho u_{\alpha} =0
\label{continuity}
\end{equation}
and the Navier-Stokes equation
\begin{equation}
 \rho\partial_t u_\alpha+\rho u_\beta \partial_\beta
u_\alpha=\partial_\beta \tau_{\alpha\beta}+\partial_\beta
\sigma_{\alpha\beta}+\frac{\rho \tau_f}{3}
(\partial_\beta((\delta_{\alpha \beta}-3\partial_\rho
P_{0})\partial_\gamma u_\gamma+\partial_\alpha
u_\beta+\partial_\beta u_\alpha).
\label{NS}
\end{equation}
The form of this equation is similar to that for a simple fluid.  However,
the details of the stress tensor reflect the additional complications of
liquid crystal hydrodynamics.  There is a symmetric contribution
\begin{eqnarray}
\sigma^{E}_{\alpha\beta} &=&-P_0 \delta_{\alpha \beta}
-\xi H_{\alpha\gamma}(Q_{\gamma\beta}+\frac{1}{3}
\delta_{\gamma\beta})-\xi (Q_{\alpha\gamma}+\frac{1}{3}
\delta_{\alpha\gamma})H_{\gamma\beta}\nonumber\\
& & \quad +2\xi
(Q_{\alpha\beta}+\frac{1}{3}\delta_{\alpha\beta})Q_{\gamma\epsilon}
H_{\gamma\epsilon}-\partial_\beta Q_{\gamma\nu} 
\frac{\delta {\cal F}}{\delta\partial_\alpha Q_{\gamma\nu}}
\label{BEstress}
\end{eqnarray}
and an antisymmetric contribution
\begin{equation}
 \tau_{\alpha \beta} = Q_{\alpha \gamma} H_{\gamma \beta} -H_{\alpha
 \gamma}Q_{\gamma \beta} .
\label{as}
\end{equation}
The hydrostatic pressure $P_0$ is defined below, while
$\eta\equiv\frac{\rho\tau_f}{3}$ is an isotropic viscosity 
($\tau_f$ is related to the details of the lattice Boltzmann algorithm,
see Eq. \ref{eq4}). If more elastic constants are 
introduced in the theory, $\tau_{\alpha \beta}$ also contains the
antisymmetric contribution from 
$\partial_\beta Q_{\gamma\nu}
\frac{\delta {\cal F}}{\delta\partial_\alpha Q_{\gamma\nu}}$.

The lattice Boltzmann algorithm described in the next subsection can be used 
for any model of the above form.  For the examples in the last section we will 
use a specific model determined by a Landau-de Gennes free energy 
(de Gennes \& Prost, 1993; Doi \& Edwards, 1989)
\begin{equation}
{\cal F}=\int d^3 r \left\{ \frac{A_0}{2}(1-\frac{\gamma}{3})
  Q_{\alpha \beta}^2 - \frac{A_0\gamma}{3} Q_{\alpha \beta}
Q_{\beta \gamma}Q_{\gamma \alpha}+ \frac{A_0\gamma}{4}
  (Q_{\alpha \beta}^2)^2 + \frac{\kappa}{2} (\partial_\alpha Q_{\beta \lambda})^2
  \right\}
\label{free}
\end{equation}
where $A_0$ is a constant and $\kappa$ denotes the elastic constant of the
liquid crystal.
We shall work within the one elastic constant approximation. Although it is 
not hard to include more general elastic terms this simplification will not 
affect the qualitative behaviour. The free energy (\ref{free}) describes a 
first order transition from the isotropic to the nematic phase,
controlled by the parameter $\gamma$.
The hydrostatic pressure $P_0$ is taken to be
\begin{equation}
P_0 = \rho T -\frac{\kappa}{2}(\nabla{\bf Q})^2,
\end{equation}
and is constant in the simulations to a very good approximation.

\subsection{Lattice Boltzmann algorithm}

\begin{figure}
\centerline{\includegraphics[width=3.8in]{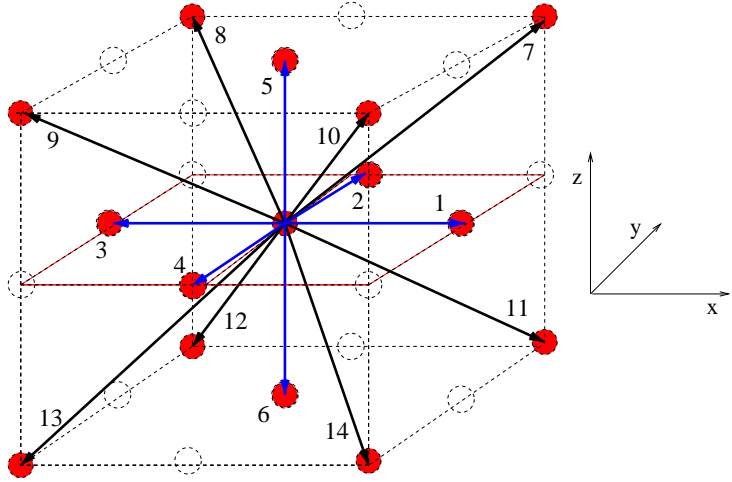}}
\caption{Lattice geometry and lattice vectors for the three dimensional 
lattice Boltzmann model.}
\label{fig1}
\end{figure}

Usually lattice Boltzmann algorithms, describing the Navier-Stokes 
equations of a simple fluid, are defined in terms of a single set of partial 
distribution functions, the scalars $f_i (\vec{x})$, that sum on each lattice
site $\vec{x}$ to give the density (Chen \& Doolen 1998).  
For liquid crystal hydrodynamics, this
must be supplemented by a second set, the symmetric traceless tensors 
${\bf G}_i (\vec{x})$, that are related to the tensor order parameter 
${\bf Q}$. Each $f_i$, ${\bf G}_i$ is associated with a lattice
vector ${\vec e}_i$.  We choose a 15-velocity model on the cubic
lattice with lattice vectors:
\begin{eqnarray}
\vec {e}_{i}^{(0)}&=& (0,0,0)\\
\vec {e}_{i}^{(1)}&=&(\pm 1,0,0),(0,\pm 1,0), (0,0,\pm 1)\\
\vec {e}_{i}^{(2)}&=&(\pm 1, \pm 1, \pm 1).
\label{latvects}
\end{eqnarray}
The indices, $i$, are ordered so that $i=0$ corresponds to 
$\vec {e}_{i}^{(0)}$, $i=1,\cdots 6$ correspond to the $\vec {e}_{i}^{(1)}$ 
set and
$i=7,14$ to the $\vec {e}_{i}^{(2)}$ set, as illustrated in Fig.~\ref{fig1}.

Physical variables are defined as moments of the distribution function:
\begin{equation}
\rho=\sum_i f_i, \qquad \rho u_\alpha = \sum_i f_i  e_{i\alpha},
\qquad {\bf Q} = \sum_i {\bf G}_i.
\label{eq1}
\end{equation}

The distribution functions evolve in a time step $\Delta t$ according
to
\begin{equation}
f_i({\vec x}+{\vec e}_i \Delta t,t+\Delta t)-f_i({\vec x},t)=
\frac{\Delta t}{2} \left[{\cal C}_{fi}({\vec x},t,\left\{f_i
\right\})+ {\cal C}_{fi}({\vec x}+{\vec e}_i \Delta
t,t+\Delta
t,\left\{f_i^*\right\})\right]
\label{eq2}
\end{equation}
\begin{equation}
{\bf G}_i({\vec x}+{\vec {e}}_i \Delta t,t+\Delta t)-{\bf G}_i({\vec
x},t)=\frac{\Delta t}{2}\left[ {\cal C}_{{\bf G}i}({\vec
x},t,\left\{{\bf G}_i \right\})+
                {\cal C}_{{\bf G}i}({\vec x}+{\vec {e}}_i \Delta
                t,t+\Delta t,\left\{{\bf G}_i^*\right\})\right]
\label{eq3}
\end{equation}
This represents free streaming with velocity ${\vec e}_i$ followed by a
collision step which allows the distribution to relax towards
equilibrium.
$f_i^*$ and ${\bf G}_i^*$ are first order approximations to
$f_i({\vec {x}}+{\vec {e}}_i \dt,t+\dt)$ and ${\bf G}_i({\vec {x}}+{\vec {e}}_i \dt,t+\dt)$
respectively. 
They are obtained by using $\Delta t\, {\cal C}_{{\bf f}i}({\vec
x},t,\left\{{\bf f}_i \right\})$ on the right hand side of Eq. (\ref{eq2}) and 
a similar substitution in Eq. (\ref{eq3}).
Discretizing in this way, which is similar to a predictor-corrector
scheme, has the advantages that lattice viscosity terms are eliminated
to second order and that the stability of the scheme is improved.

The collision operators are taken to have the form of a single
relaxation time Boltzmann equation, together with a forcing term
\begin{equation}
{\cal C}_{fi}({\vec {x}},t,\left\{f_i \right\})=
-\frac{1}{\tau_f}(f_i({\vec {x}},t)-f_i^{eq}({\vec {x}},t,\left\{f_i
\right\}))
+p_i({\vec {x}},t,\left\{f_i \right\}),
\label{eq4}
\end{equation}
\begin{equation}
{\cal C}_{{\bf G}i}({\vec x},t,\left\{{\bf G}_i
\right\})=-\frac{1}{\tau_{\bf G}}({\bf G}_i({\vec x},t)-{\bf
G}_i^{eq}({\vec x},t,\left\{{\bf G}_i \right\}))
+{\bf M}_i({\vec x},t,\left\{{\bf G}_i \right\}).
\label{eq5}
\end{equation}

The form of the equations of motion and thermodynamic equilibrium
follow from the choice of the moments of the equilibrium distributions
$f^{eq}_i$ and ${\bf G}^{eq}_i$ and the driving terms $p_i$ and
${\bf M}_i$. $f_i^{eq}$ is constrained by
\begin{equation}
\sum_i f_i^{eq} = \rho,\qquad \sum_i f_i^{eq} e_{i \alpha} = \rho
u_{\alpha}, \qquad
\sum_i f_i^{eq} e_{i\alpha}e_{i\beta} = \Pab+\rho
u_\alpha u_\beta
\label{eq6}
\end{equation}
where the zeroth and first moments are chosen to impose conservation of
mass and momentum. The second moment of $f^{eq}$ controls the symmetric
part of the stress tensor, whereas the moments of $p_i$
\begin{equation}
\sum_i p_i = 0, \quad \sum_i p_i e_{i\alpha} = \partial_\beta
\tau_{\alpha\beta},\quad \sum_i p_i
e_{i\alpha}e_{i\beta} = 0
\label{eq7}
\end{equation}
impose the antisymmetric part of the stress tensor.
For the equilibrium of the order parameter distribution we choose
\begin{equation}
\sum_i {\bf G}_i^{eq} = {\bf Q},\qquad \sum_i
{\bf G}_i^{eq} {e_{i\alpha}} = {\bf Q}{u_{\alpha}},
\qquad \sum_i {\bf G}_i^{eq}
e_{i\alpha}e_{i\beta} = {\bf Q} u_\alpha u_\beta .
\label{eq8}
\end{equation}
This ensures that the order parameter
is convected with the flow. Finally the evolution of the
order parameter is most conveniently modeled by choosing
\begin{equation}
\sum_i {\bf M}_i = \Gamma {\bf H}({\bf Q})
+{\bf S}({\bf W},{\bf Q}) \equiv {\bf \hat{H}}, \qquad
\qquad \sum_i {\bf M}_i {e_{i\alpha}} = (\sum_i {\bf M}_i)
{u_{\alpha}}
\label{eq9}
\end{equation}
which ensures that the fluid minimises its free energy at equilibrium.
Conditions (\ref{eq6})--(\ref{eq9}) are satisfied, as is usual in lattice 
Boltzmann schemes, by writing the equilibrium distribution functions and 
forcing terms as polynomial expansions in the velocity
\begin{eqnarray}
f_i^{eq}&=&A_s + B_s u_\alpha e_{i\alpha}+C_s u^2+D_s u_\alpha u_\beta
e_{i\alpha}e_{i\beta}+E_{s\alpha\beta}e_{i\alpha}e_{i\beta},\nonumber \\
{\bf G}_i^{eq}&=&{\bf J}_s + {\bf K}_s u_\alpha e_{i\alpha}+{\bf L}_s
u^2+{\bf N}_s u_\alpha
u_\beta e_{i\alpha}e_{i\beta},\nonumber \\
p_i&=&T_s \partial_\beta \tau_{\alpha\beta} e_{i\alpha},\nonumber \\
{\bf M}_i&=&{\bf R}_s+{\bf S}_s u_\alpha e_{i\alpha},
\end{eqnarray}
where $s \in \{0,1,2\}$ identifies separate coefficients for
the velocity vectors $\vec {e}_{i}^{(s)}$.  The coefficients are determined
by evaluating the constraints Eqs. (\ref{eq6})--(\ref{eq9}) and matching terms.
In doing this, we have made use of the following
symmetry relations for the lattice vectors, Eq.(\ref{latvects}), 
\begin{eqnarray}
&&\sum_{i=1}^{6}e_{i\alpha}^{(1)} =
\sum_{i=7}^{14}e_{i\alpha}^{(2)} =0
\qquad \alpha=1,2,3 \\
&&\sum_{i=1}^{6}e_{i\alpha}^{(1)}e_{i\beta}^{(1)} = 2
\delta_{\alpha\beta} \qquad
\sum_{i=7}^{14}e_{i\alpha}^{(2)}e_{i\beta}^{(2)}=8\delta_{\alpha\beta} \\
&&\sum_{i=1}^{6}e_{i\alpha}^{(1)}e_{i\beta}^{(1)}e_{i\eta}^{(1)} =
  \sum_{i=7}^{14}e_{i\alpha}^{(2)}e_{i\beta}^{(2)}e_{i\eta}^{(2)}=0\\
&&\sum_{i=1}^{6}
e_{i\alpha}^{(1)}e_{i\beta}^{(1)}e_{i\eta}^{(1)}e_{i\zeta}^{(1)}
=2\delta_{\alpha\beta}\delta_{\beta\eta}\delta_{\eta\zeta}\\
&& \sum_{i=7}^{14}
e_{i\alpha}^{(2)}e_{i\beta}^{(2)}e_{i\eta}^{(2)}e_{i\zeta}^{(2)}
=8\Delta_{\alpha\beta\eta\zeta}-16\delta_{\alpha\beta}\delta_{\beta\eta}\delta_{\eta\zeta}.
\end{eqnarray}
where $\Delta_{\alpha\beta\eta\zeta}=\delta_{\alpha\beta}\delta_{\eta\zeta}
+\delta_{\alpha\eta}\delta_{\beta\zeta}+\delta_{\alpha\zeta}\delta_{\beta\eta}$.

The expansion coefficients are then determined to be:
\begin{eqnarray}
&&A_2=\frac{1}{10}(Tr{\bf P}/3),\qquad A_1= A_2,\qquad A_0=\rho-14 A_2,\nonumber \\
&&B_2=\rho/24,\qquad B_1= 8B_2,\nonumber \\
&&C_2=-\frac{\rho}{24},\qquad C_1=2C_2,
\qquad C_0=-\frac{2}{3}\rho,\nonumber \\
&&D_2=\frac{\rho}{16}, \qquad D_1=8D_2\nonumber \\
&&E_{2\alpha\beta}=\frac{1}{16}(\Pab -Tr{\bf P}/3
\delta_{\alpha\beta}),\quad
E_{1\alpha\beta}= 8E_{2\alpha\beta},\nonumber \\
&&{\bf J}_0={\bf Q},\qquad {\bf K}_2={\bf Q}/24,\qquad {\bf K}_1= 8{\bf K}_2,\nonumber \\
&&{\bf L}_2=-\frac{{\bf Q}}{24},\qquad {\bf L}_1=2{\bf L_2}, \qquad
{\bf L}_0=-\frac{2 {\bf Q}}{3},\nonumber \\
&&{\bf N}_2=\frac{{\bf Q}}{16}, \qquad
{\bf N}_1=8{\bf N}_2\nonumber \\
&&{\bf R}_2=\widehat{\bf H}/15, \qquad
{\bf R}_1={\bf R}_0={\bf R}_2\nonumber \\
&&{\bf S}_2=\frac{\widehat{\bf H}}{24}, \qquad {\bf S}_1=8 {\bf S}_2,\nonumber \\
&&T_2=1/24, \qquad T_1=8 T_2. \label{coeff}
\end{eqnarray}

\subsection{Velocity boundary conditions}
To illustrate the implementation of the boundary conditions for a system sheared along $y$ consider a wall at $z=0$. No flux across the wall at $z=0$ implies 
$\sum_i f_i e_{iz} = 0$ and hence
\begin{equation}
f_5+f_7+f_8+f_9+f_{10}=f_6+f_{11}+f_{12}+f_{13}+f_{14}.
\label{noflux}
\end{equation}
For no slip along the $x$-direction $\sum_i f_i e_{ix} = 0 $ or
\begin{equation}
f_1+f_7+f_{10}+f_{11}+f_{14} =
f_3+f_8+f_9+f_{12}+f_{13}.
\label{noslipx}
\end{equation}
For fixed velocity $u_y^*$ along the y direction 
$\sum_i f_i e_{iy} =  \rho u_y^*$
or
\begin{equation}
f_2+f_7+f_8+f_{11}+f_{12}-f_4-f_9-f_{10}-f_{13}-f_{14}=\rho u_y^*.
\label{noslipy}
\end{equation}
For the wall at $z=0$ there are $5$ unknown distributions 
$f_5,f_7,f_8,f_9$ and $f_{10}$. To determine these, two further constraints are required in addition to the relations 
(\ref{noflux}), (\ref{noslipx}), and (\ref{noslipy}). 
Symmetry suggests
\begin{equation}
f_{7}-f_{8} = f_{10} - f_{9}.
\label{symm}
\end{equation}
The conservation of mass is easily obtained by choosing
\begin{equation}
f_5 = f_6.
\label{bb}
\end{equation}
Other choices can be made for these two further constraints. With
the present one, 
equations (\ref{noflux})-(\ref{bb}) can now be solved to give:
\begin{eqnarray}
f_5 &=& f_6, \nonumber \\
f_{7}&=& \frac{1}{4}\left( -f_1-f_2+f_3+f_4-f_{11}+f_{12}+3f_{13}+f_{14} +\rho
u_y^*\right),\nonumber \\
f_{8}&=& \frac{1}{4}\left( f_1-f_2-f_3+f_4+f_{11}-f_{12}+f_{13}+3f_{14} +\rho
u_y^*\right),\nonumber \\
f_{9}&=& \frac{1}{4}\left( f_1+f_2-f_3-f_4+3f_{11}+f_{12}-f_{13}+f_{14} -\rho
u_y^*\right),\nonumber \\
f_{10}&=& \frac{1}{4}\left( -f_1+f_2+f_3-f_4+f_{11}+3f_{12}+f_{13}-f_{14} -\rho
u_y^*\right).
\end{eqnarray}

\section{Numerical Results}

Our main purpose here is to present in detail a
numerical algorithm for simulating liquid crystal hydrodynamics
in three dimensions. Consequently, 
we restrict ourselves to presenting a few simple
examples, aimed at checking the approach.  Further
numerical applications are listed in the 
concluding section and will be reported elsewhere.

To apply the algorithm  to
systems of practical relevance, we need to
map the simulation parameters listed in Section (2b)
onto physical numbers which can be compared to
real experiments. This task is not trivial and proceeds as
follows. First, we have to fix a physical value for the
energy scale, which in simulations is given by $\rho T$.
Second, the size of the simulation lattice along $z$, 
$L_z$, is matched to the liquid crystal cell thickness.
Third, the parameter $\Gamma$ is related to the magnitude of
order and the physical value of the Miesowicz viscosity $\gamma_1$
(see De Gennes \& Prost 1993) via the formula
$\Gamma=2q^2/\gamma_1$ (Denniston et al. 2001).
This fixes the time step in the simulation.
Once the energy, length and time scales are specified, all
simulation parameters can be related straightforwardly to
experimental quantities.

As specific examples of applications of the three-dimensional
algorithm we propose two hydrodynamic effects in two commonly used
twisted nematic cells.
The simplest example of such cells is obtained when at the two surfaces
(perpendicular to the z-axis, say)  the director is anchored and
the anchoring is homogeneous, i.e. the directors lie in the surface plane. 
If the two anchoring
directions at $z=0$ and $z=L_z$ are mutually perpendicular, 
the free energy minimum
in the absence of other factors is a twisted director field along
the $z$ direction. This is the geometry we consider (with the
two pinning directions making an angle of $\pm\pi/4$  with the $x$ axis to maximise
the system symmetry). 

In devices, this cell is switched on
by means of an electric field perpendicular to the surfaces. 
Here we shall study the dynamics when the
cell is switched off and relaxes back to the twisted state. Therefore
we initialize the director field to lie along the $z$ direction in the whole
cell apart from two small regions (8 lattice sites in our simulations) 
near the top and bottom boundaries to allow a smooth relaxation between surface and bulk order.
The   numerical parameters  used  for this calculation were as follows:
$\kappa=0.05$, $\gamma=3.5$,  $L_z=91$, $\Gamma=0.33775$, $\xi=0.8$     
and $\tau_f=0.56$.  By means of the procedure outlined above,
the simulation system can be mapped
onto a physical device of thickness $4.5$  $\mu m$, with all three
Frank elastic  constants  equal  to $\sim 9.3$ pN, while
$\Gamma=0.58$ Poise$^{-1}$ and the
Miesowicz coefficient $\gamma_1=1.3$ Poise.

\begin{figure}
\centerline{\includegraphics[angle=270,width=3.5in]{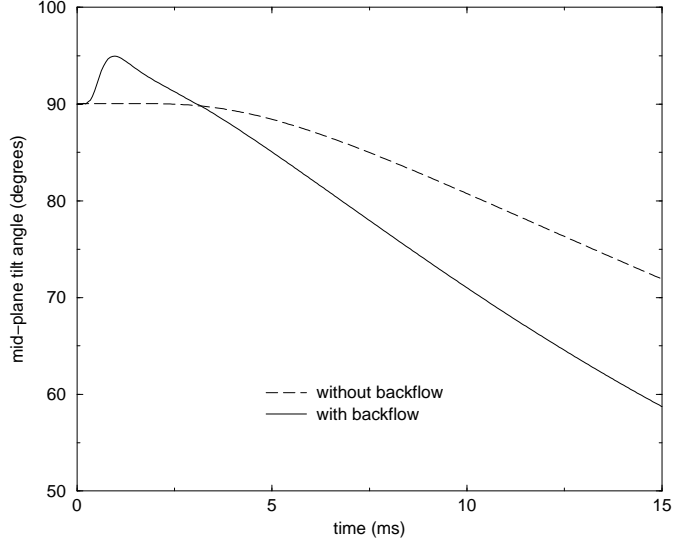}}
\caption{Evolution of the tilt of the director field at the centre of a 
twisted nematic cell after the field is switched off.
Notice that $\theta$ and $\pi-\theta$ actually represent the same state
since the director field has no 'head' or 'tail'.}
\label{fig2}
\end{figure}

In Fig. \ref{fig2} we show how the tilt, $\theta$, of the director in the 
midplane ($z=L_z/2$) changes as  a function of  time  immediately after the device is switched off. 
(The tilt angle is the angle with respect to the $xy$ plane).
In the
figure we compare the results obtained when backflow is neglected 
and when it is included.  
It  can be seen  that backflow has a significant effect on the switching,  
namely it causes the non-monotonicity 
in the relaxation  dynamics of the mid-plane tilt angle.
The flow in  the device causes  the   director to  start  its
relaxation in  the 'wrong'  direction. As the system is
symmetric the mid-plane tilt has to pass  through $90^{\circ}$ again
during the relaxation to the twisted state.
This is a known effect which is called of optical bounce.
Note also that with
backflow the relaxation is faster.

Optical bounce has only recently been observed directly in an experiment
(Smith et al. 2002, Ruan et al. 2002). 
Our curve compares qualitatively well with the one observed  
by Ruan et al., who use a cell of comparable thickness
and $\gamma_1$ value, but with a slightly larger value of the
elastic constants.
In the experiments, the maximum in the tilt angle occurred
for a value of $\theta\sim 96^{\circ}$, the backflow effect lasted $\sim 2$
ms, and after $10$ ms the angle had relaxed down to $\sim$ $60^{\circ}$.
Since the details of the dynamics depend  on the 
width of the region which is initially oriented along $z$ and on the 
temperature, we are satisfied with  semiquantitative agreement here.  

The optical bounce can be seen within Leslie-Ericksen 
theory (see e.g. Kelly et al. 1999).
However with the Beris-Edwards equations solved here we can also investigate
the dependence of
the dynamics on temperature. The bounce gets deeper as the
effective temperature is decreased for constant $\kappa$, or if $\kappa$ is increased at constant $\gamma$. Interestingly, as the temperature increases and approaches the
isotropic-to-nematic transition point, the waiting time needed for the 
system to start the bounce increases. This happens because the magnitude  of the order near the surface, 
where the initial director deformations are confined, drops significantly allowing 
two defects to transiently form. These defects block the system
evolution and slow down the start-up of the dynamics. 
Furthermore, we can assess the effect of
tuning the strength of anchoring and the magnitude of surface order on the
bounce magnitude: while the former has a negligible effect on the
result, the latter can have a bigger impact, i.e. if the order 
within the surface is slightly different from the typical order within the
bulk, the bounce magnitude can change significantly.



As a second example, we consider another twisted geometry, in which the director
is anchored parallel on the two boundaries, and a $\pi$ twist is introduced in the $z$ direction. 
We consider a shear experiment in which the bounding plates move at constant velocities $\pm u_y$.
Because the director does not lie in the shear plane, (the
$yz$ plane), the hydrodynamics is complex. In particular,
experiments and approximate modelling (using the Leslie-Ericksen equations)
predict that even though the system is sheared along $y$, a secondary flow
appears along the $x$ direction (De Gennes \& Prost 1993).

\begin{figure}
\centerline{\includegraphics[width=6.in]{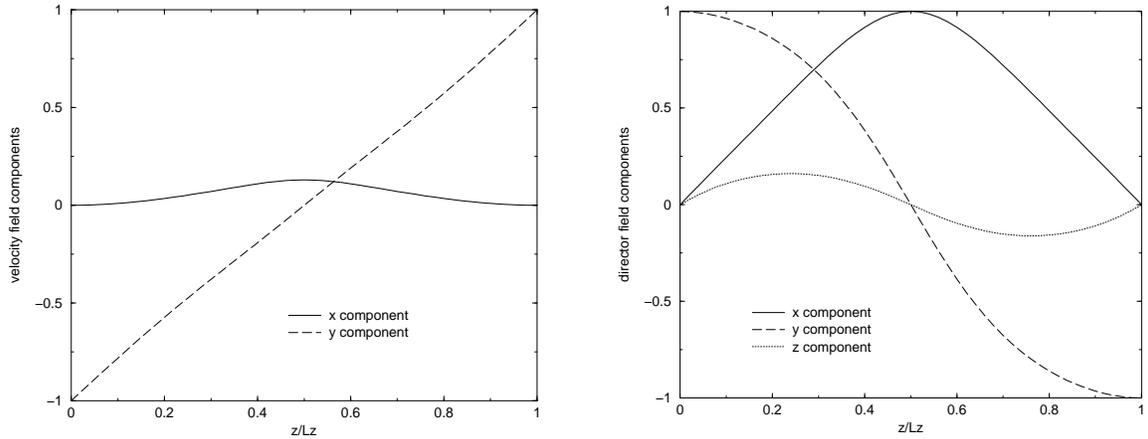}}
\caption{Velocity field (left) and director field (right) as a 
function of the position across the cell  for
a $\pi$ twisted nematic liquid crystal under shear.
Note the secondary flow along $x$ in the left figure.
The velocity profile in the left plot is scaled by the shearing velocity, 
which was $\sim 0.8$ mm/s. }
\label{fig3}
\end{figure}

In Figure \ref{fig3}, we show that the lattice Boltzmann simulations
can reproduce this three-dimensional feature well. 
In the same figure, we also show
how the director field components $n_x$, $n_y$ and $n_z$ vary across the cell.
These results were obtained for simulation parameters 
$\kappa=0.05$, $\gamma=3.5$, $L_z=51$, $\Gamma=0.33775$, $\xi=0.85$     
and $\tau_f=0.56$, which can be mapped this time 
onto a physical device of thickness $2.25$  $\mu m$, with all three
Frank elastic  constants  equal  to $\sim 3.3$ pN.

\section{Discussion}

In this paper we have described in detail a lattice Boltzmann
algorithm to simulate liquid crystal hydrodynamics in three dimensions.  
We have shown that it can, for values of the simulation parameters which 
capture physical nematic properties and device dimensions, reproduce optical 
bounce in a twisted nematic device and secondary flow in a twisted nematic 
under shear. Future applications will include studying backflow in more 
complicated device geometries such a pi-cell (Jung et al. 2003) and 4-domain 
twisted nematic cells. It will also be of interest to consider the flow of 
liquid crystals in narrow channels as, with the advent of microchannel 
technology, quantitative experiments may soon become possible. 
Here, surface anchoring and elastic deformations of the 
director field are expected to heavily impact on the
rheological properties (Marenduzzo et al. 2003).
Other areas where this approach will be useful is in the study of shear 
banding under both constant stress and constant strain and in considering 
the flow of cholesterics, nematic liquid crystals with a 
director field that form a helix.


\begin{thebibliography}{}

\item 
P.G. de Gennes \& J. Prost 1993, {\it The Physics of
Liquid Crystals, 2nd Ed.}, Clarendon Press, Oxford, (1993).

\item
A.N. Beris \& B.J. Edwards 1994, {\it Thermodynamics of Flowing Systems},
Oxford: Oxford University Press. 

\item
A. D. Rey \& M. M. Denn 2002, Dynamical phenomena in liquid-crystalline
materials, {\it Annu. Rev. Fluid Mech.} {\bf 34}, 233-266.

\item
A.N. Beris, B.J. Edwards \& M. Grmela 1990, 
Generalized constitutive equation for polymeric liquid-crystals. 1.
Model formulation using the Hamiltonian (Poisson bracket) formulation
{\it J. Non-Newton. Fluid Mechanics}, {\bf 35} 51-72.

\item
S. Chen \& G. D. Doolen 1998,
Lattice Boltzmann method for fluid flows,
{\it Annu. Rev. Fluid Mech.} {\bf 30}, 329-364.

\item
S. Succi 2001,
{\it The lattice Boltzmann equation for fluid dynamics
and beyond}, Oxford University Press.

\item
D. A. Wolf-Gladrow 2000, {\it Lattice-gas cellular automata
and lattice Boltzmann models}, Vol. 1725, Springer, Berlin.

\item
R. Benzi, S. Succi, M. Vergassola 1992,
The lattice Boltzmann equation: theory and applications,
{\it Phys. Rep.} {\bf 222}, 145-197.

\item
M. Doi \& S. F. Edwards 1989, {\it The Theory of Polymer Dynamics}, 
Clarendon Press, Oxford.

\item
C. Denniston, E. Orlandini \& J. M. Yeomans 2000,
Simulations of liquid crystal hydrodynamics in the isotropic and 
nematic phases,
{\it Europhys. Lett.} {\bf 52}, 481-487.

\item
C. Denniston, E. Orlandini \& J. M. Yeomans 2001,
Lattice Boltzmann simulations of liquid crystal hydrodynamics,
{\it Phys. Rev. E} {\bf 63}, 056702.

\item
N. J. Smith, M. D. Tillin \& J. R. Sambles 2002, 
Direct optical quantification of backflow in a 90 degrees twisted nematic 
cell, {\it Phys. Rev. Lett.} {\bf 88}, 088301.

\item
L. Z. Ruan \& J. R. Sambles 2002, 
Dynamics of a twisted nematic cell using a convergent beam system,
{\it J. Appl. Phys.} {\bf 92}, 4857-4862.

\item J. Kelly, S. Jamal \& M. Cui 1999,
Simulation of the dynamics of twisted nematic devices
including flow, {\it J. Appl. Phys.} {\bf 86}, 4091-4095.

\item 
J. Jung, C. Denniston, E. Orlandini \& J. M. Yeomans 2003,
Anisotropy of domain growth in nematic liquid crystals,
{\it Liquid Crystals} in press.

\item 
D. Marenduzzo, E. Orlandini \& J. M. Yeomans 2003,
Rheology of distorted nematic liquid crystals,
{\it Europhys. Lett.}, {\bf 64}, 406-412.
















\end{thebibliography}
\end{document}